\documentclass[final,twocolumn,aps,floats,amsfonts,amssymb,showpacs,showkey]{revtex4}
\usepackage{amsmath}
\usepackage{graphicx}
\usepackage{color}
\usepackage{hyperref}
\usepackage{psfrag}

\begin{document}

\title{Characterization of turbulence in inhomogeneous anisotropic flows}

\author{Pierre-Philippe Cortet}
\author{Pantxo Diribarne}
\author{Romain Monchaux}
\author{Arnaud Chiffaudel}
\author{Fran\c cois Daviaud}
\author{B\'ereng\`ere Dubrulle}
\email{berengere.dubrulle@cea.fr}

\affiliation{Service de Physique de l'\'Etat Condens\'e, DSM, CEA
Saclay, CNRS URA 2464, 91191 Gif-sur-Yvette, France}

\date{\today}


\begin{abstract}
We introduce a global quantity $\delta$ that characterizes
turbulent fluctuations in inhomogeneous anisotropic flows. This
time-dependent quantity is based on spatial averages of global
velocity fields rather than classical temporal averages of local
velocities. $\delta(t)$ provides a useful quantitative
characterization of any turbulent flow through generally only two
parameters, its time average $\bar \delta$ and its variance
$\delta_2$. Properties of $\bar \delta$ and $\delta_2$ are
experimentally studied in the typical case of the von K\'arm\'an
flow and used to characterize the scale by scale energy budget as
a function of the forcing mode as well as the transition between
two flow topologies.
\end{abstract}

\maketitle

\section{Introduction}
\begin{figure*}
\psfrag{r/R}[c][][1]{$r/R$}\psfrag{z/Z}[c][][1.1]{$z/Z$}
\begin{tabular}{ccc}
(a)&(b)&(c)\\
   \includegraphics[width=.77\columnwidth]{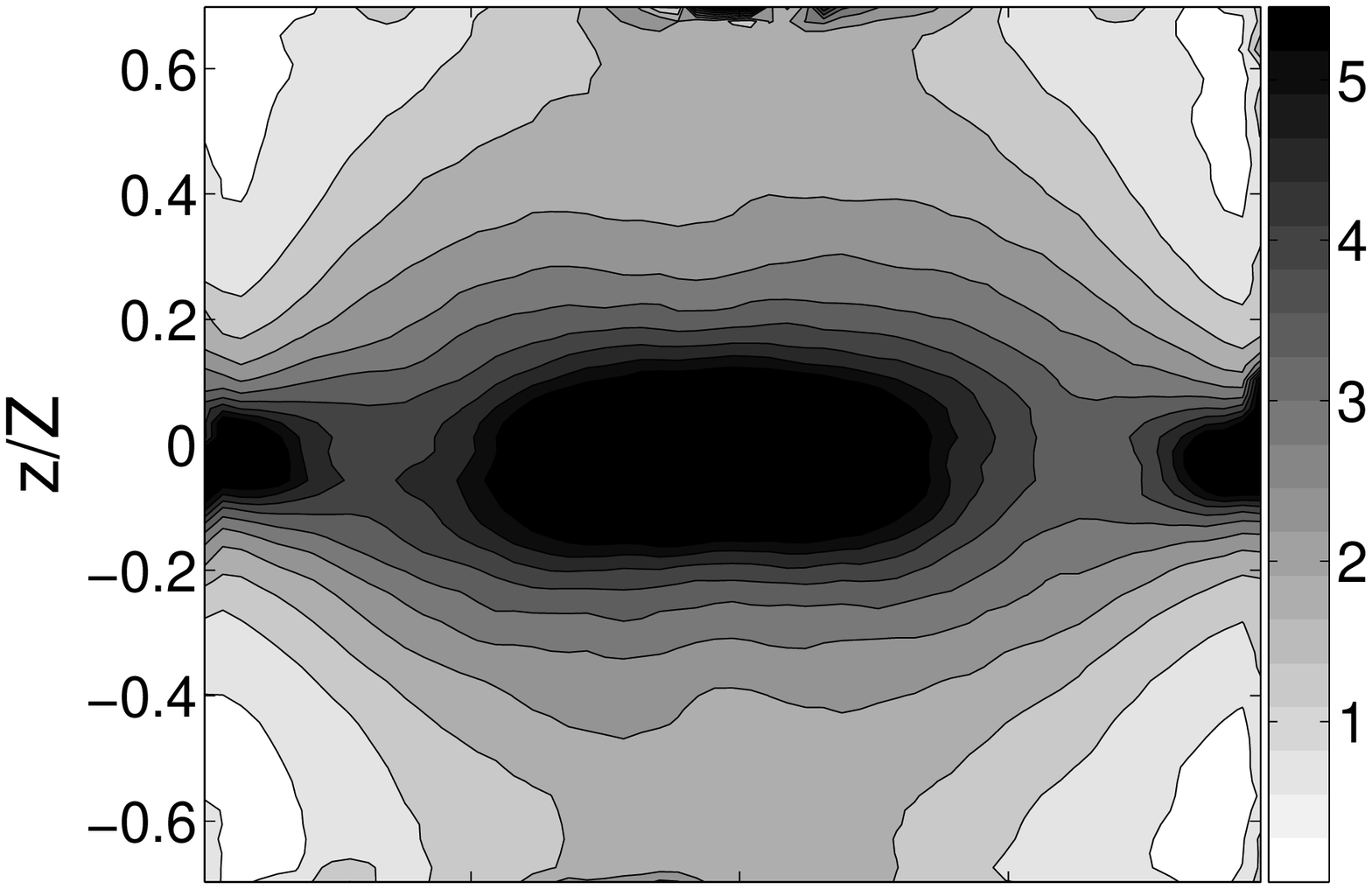} &
   \includegraphics[width=.60\columnwidth]{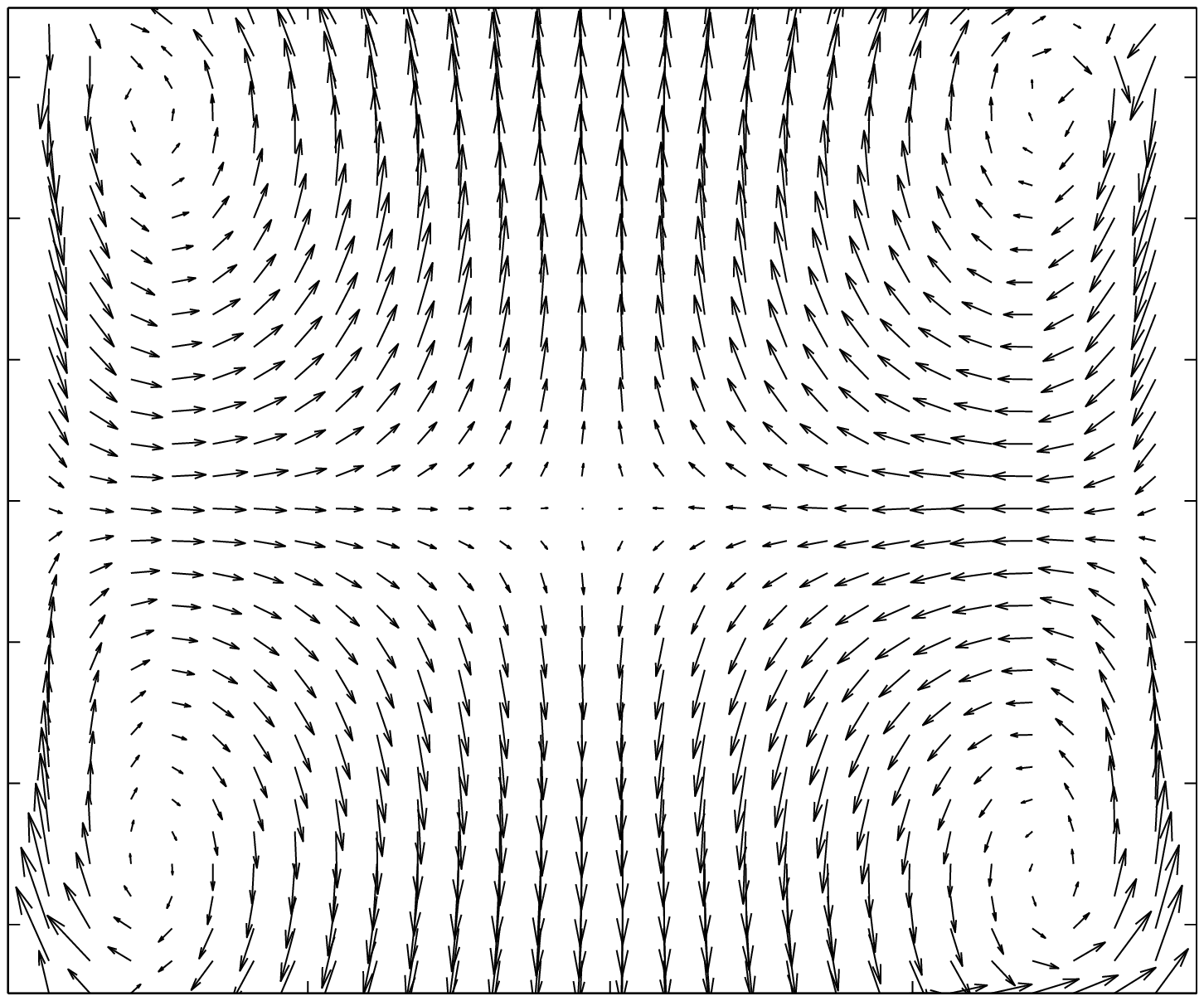} &
   \includegraphics[width=.60\columnwidth]{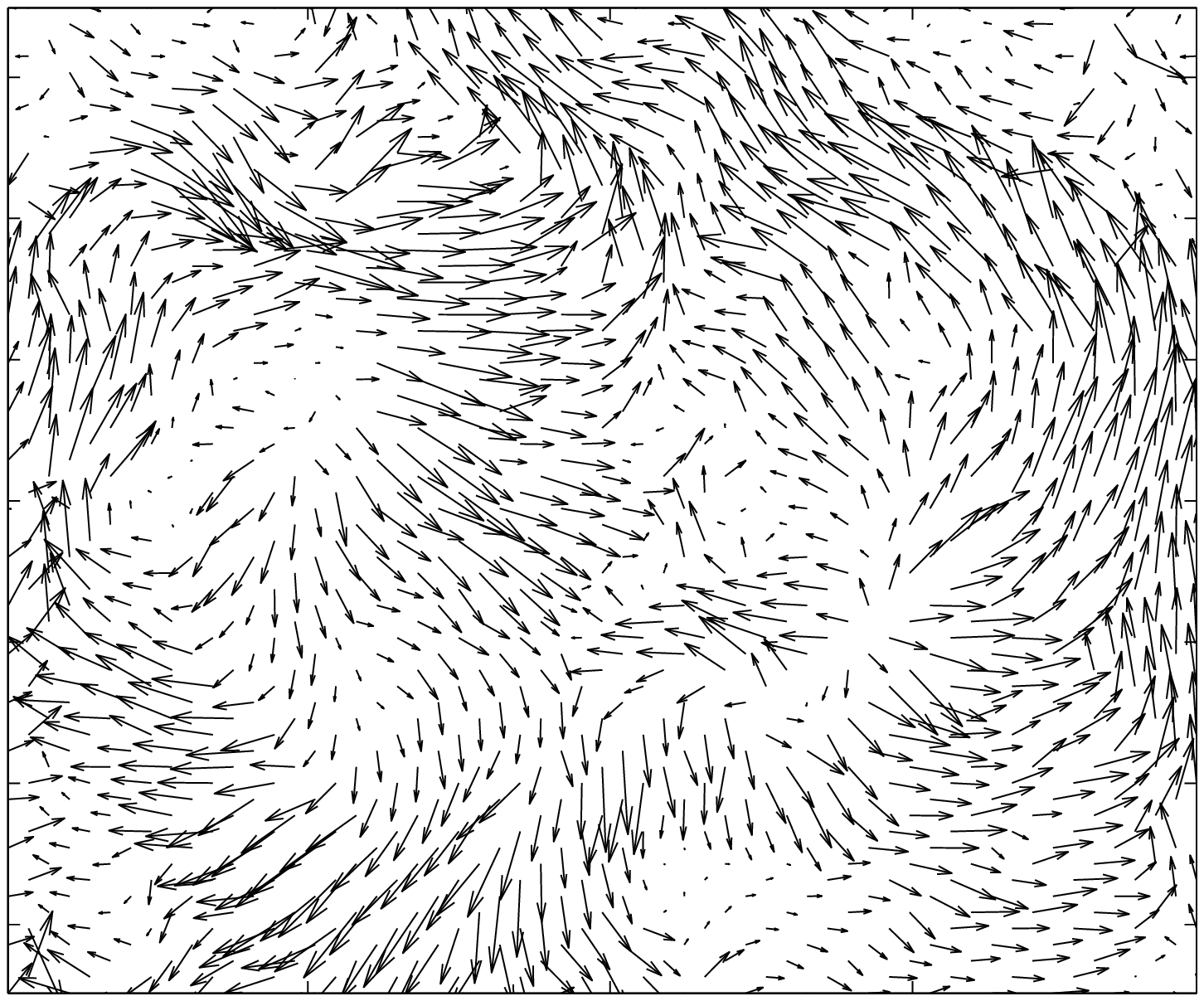} \\
   (d)&(e)&(f)\\
   \includegraphics[width=.77\columnwidth]{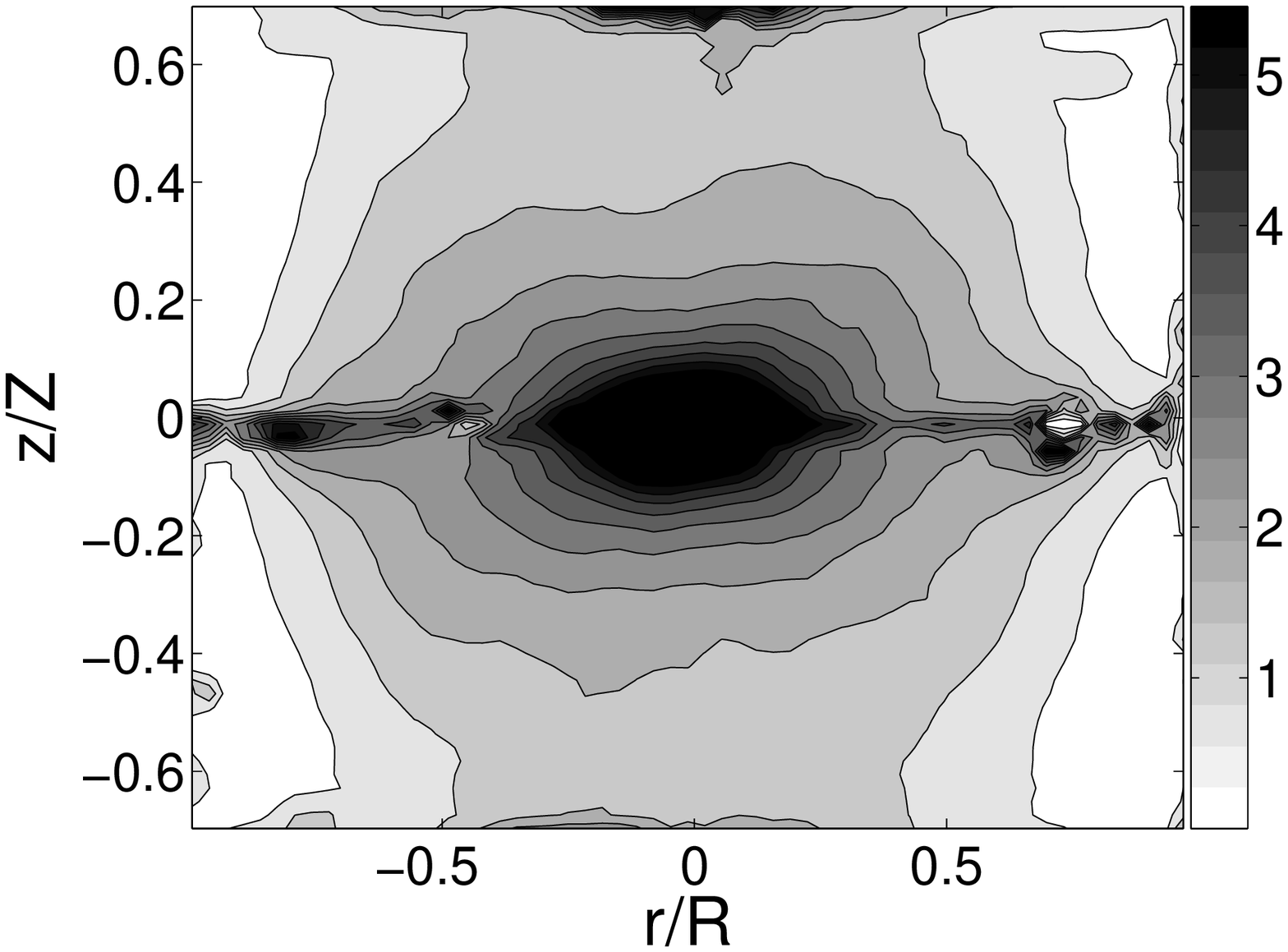} &
   \includegraphics[width=.60\columnwidth]{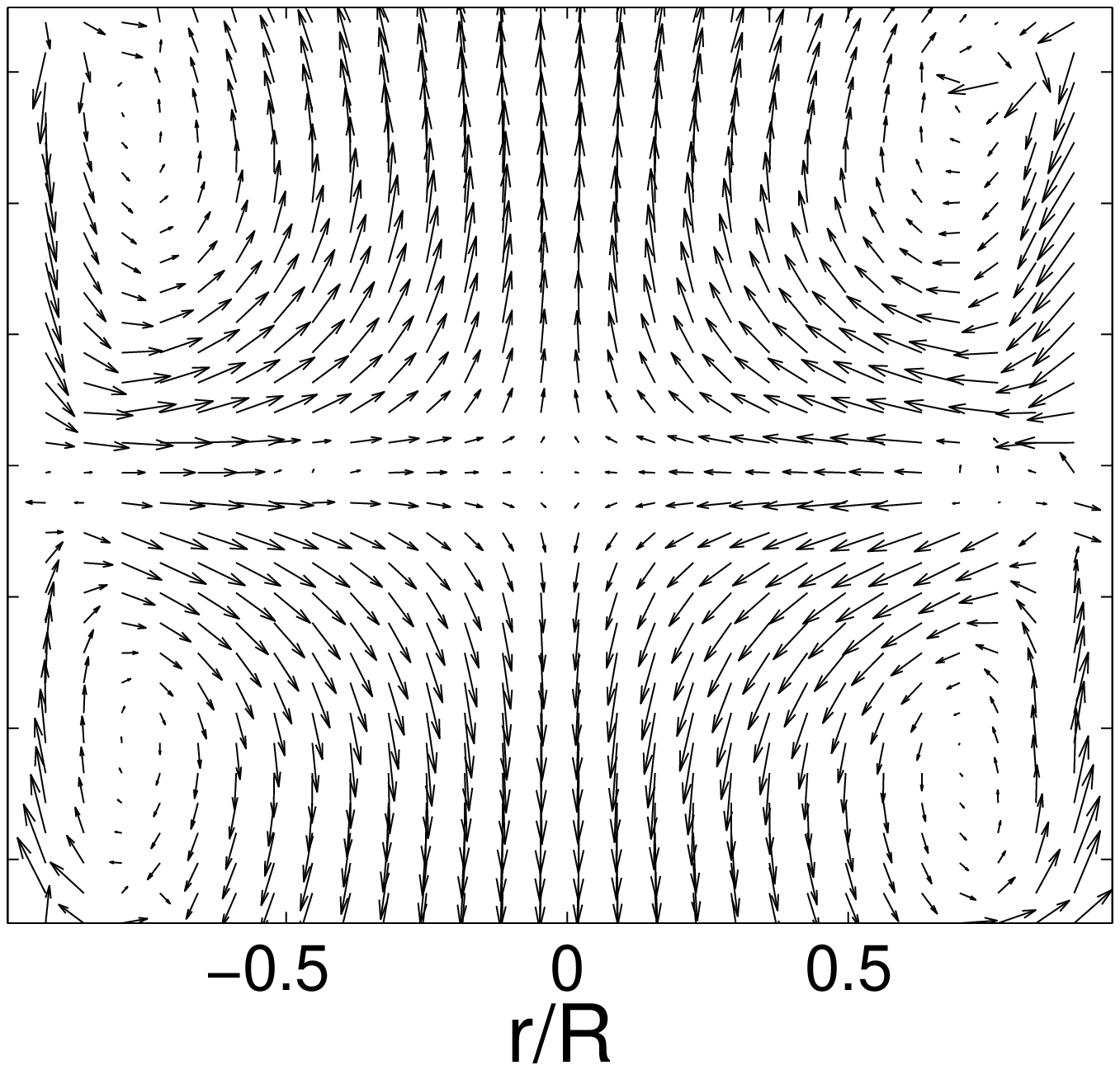} &
   \includegraphics[width=.60\columnwidth]{figure1e.eps}
  \end{tabular}
\caption{Maps of turbulence intensity $i$, (a) and (d),
corresponding time averaged, (b) and (e), and typical
instantaneous, (c) and (f), velocity fields for two von K\'arm\'an
flows ((a-c) and (d-f)) at the same Reynolds number but with
slightly different geometry. Both flows correspond to the same
experimental setup
---TM60$(+)$---, except for a narrow annulus inserted in the
median plane of the flow in (d-f). In (a) and (d), for better
visibility, the gray scale has been saturated at $i=6$ since
values up to $i\sim$~10$^4$ can be reached in the median plane due
to weak mean velocities and large fluctuations. Additionally,
vectors of the instantaneous velocities have been scaled by half
with respect to the mean field vectors.}\label{Map}
\end{figure*}

The classical theory of homogeneous isotropic turbulence applies
to  flows with zero velocity average so that all the turbulence
properties are contained in the temporal velocity fluctuations
\cite{Frisch}. Real turbulent flows, with non trivial boundary
conditions and external forcing, often possess a non-zero mean
flow that may or may not be stationary \footnote{In most of the
early experimental works about turbulence, this non-zero mean flow
is actually used to transform ``local'' temporal information into
``extended'' spatial information through the so-called Taylor
hypothesis.}. Quantifying the influence of turbulent fluctuations
on the flow dynamics has been the corner stone of turbulence
theory. In early experimental works, velocity measurements were
performed through hot wire probes providing temporal variations of
the velocity at fixed points. A natural parameter quantifying the
intensity of turbulence has therefore been introduced as: $$i
=\sqrt{\frac{\overline{V^2}-\overline{V}^2}{ \overline{V}^2}},$$
where $V$ is the velocity and $\overline X$ refers to the time
average of variable $X$. While such a parameter clearly
characterizes fluctuations in a homogeneous flow, one may question
its relevance in more general anisotropic inhomogeneous flows
where turbulence intensity depends on the measurement point (cf.
Figs. \ref{Map}(a) and (d)). For example, $i$ diverges around
stagnation points or shears so that no robust global quantity can
be built by integrating or averaging $i$ over the whole flow. This
question is receiving an increasing interest since the advent of
Particle Image Velocimetry (PIV) measurements which provide
instantaneous snapshots of the velocity field.
An illustration of the importance of that issue may be given in
the typical inhomogeneous anisotropic case of the von K\'arm\'an
flow. This flow, generated in between two counter-rotating
co-axial impellers, has received a lot of interest
\cite{cadot95,labbe96a,mordant97a,ravelet2004,nore05} as a simple
way to obtain experimentally a very large Reynolds number flow in
a compact design ($Re\sim 10^6$ in a table top water apparatus).
In the equatorial shear layer of such a flow, fluctuations are
large and exhibit similar local properties as in large Reynolds
number experimental facilities devoted to homogeneous turbulence
\cite{Maurer1994,Tabeling1996,labbe96a,Arneodo1996}. Away from the
shear layer, one observes a decrease of the turbulence intensity
\cite{theselouis,marie04} as seen in Figs. \ref{Map}(a) and (d).
Overall, the flow is strongly turbulent, so that the instantaneous
velocity fields, measured by means of a PIV system, strongly
differ in a non-trivial manner from their time average (cf. Figs.
\ref{Map}(b) and (c)).

First, we introduce in Sec. \ref{theor} a global time-dependent
quantity $\delta(t)$ that allows to characterize globally
turbulent fluctuations in any inhomogeneous anisotropic flow. In
Sec. \ref{secexp}, we present our experimental setup and show how
to compute practically $\delta(t)$ from Stereoscopic Particule
Image Velocimetry (SPIV) measurements. In Sec. \ref{appli}, we
first show that $\delta(t)$ provides a useful quantitative
characterization of turbulent flows through generally only two
parameters, its time average $\bar \delta$ and its variance
$\delta_2$, that quantify respectively the level of fluctuations
compared to the mean flow and their ability to disturb the mean
flow. These two quantities are introduced as a generalization of
the classical local turbulence intensity $i$. Finally, we discuss
applications of these new global measures of turbulence intensity:
we show how these parameters may be used to study the influence of
the forcing conditions on the flow.

\section{Generalization of the turbulence intensity}
\label{theor}

\begin{figure}
\psfrag{d}[c][][1.2]{$\delta$}
\begin{center}
\includegraphics[width=\columnwidth ]{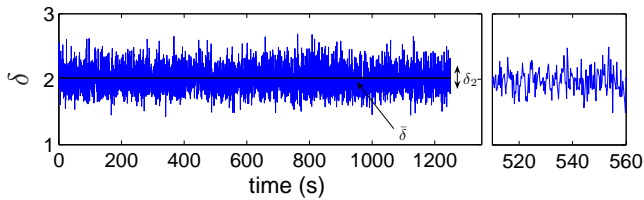}
\caption{Example of the temporal evolution of $\delta(t)$ at two
time scales: a full record of about 20 minutes on the left, and a
focus on a 40 seconds sample on the right. A schematic
illustration of the meaning of $\bar\delta$ and $\delta_2$ is also
provided over the longest time record.} \label{noise}
\end{center}
\end{figure}

We consider a general anisotropic inhomogeneous turbulent flow,
with velocity field $V(\vec x,t)=[u(\vec x,t),v(\vec x,t),w(\vec
x,t)]$. In order to smooth out the possible spatial
inhomogeneities of the turbulence intensity, we define a scalar
quantity through~:
$$ \delta(t)=\frac{\langle V^2 (t)\rangle}{\langle  \overline{V}^2 \rangle},$$
where $\langle X \rangle$ and $\overline{X}$ refer respectively to
spatial and time average of $X$, and  $V(\vec x,t)^2=u^2+v^2+w^2$
is the local kinetic energy density at time $t$. The
quantity $\delta$ can also be viewed as the ratio of the total
kinetic energy of the instantaneous flow to the total kinetic
energy of the mean flow:
\begin{equation}
\delta(t)=\frac{E\left(V(t) \right)}{E\left( \overline{V}
\right)}.
\label{delta-def}
\end{equation}
Note that this scalar parameter is time dependent and generally
widely fluctuates in time (cf. Fig.~\ref{noise}). We then define
two time independent parameters, $\bar\delta$ and $\delta_2$, that
are respectively the time average and the variance of $\delta(t)$
as:
$$\bar\delta=\overline{\langle V^2(t) \rangle}/{\langle \overline{V}^2 \rangle},$$
$$\delta_2=\sqrt{\overline{\delta(t)^2}-\overline{\delta(t)}^2}.$$ We show
in Sec.~\ref{secexp} that these two quantities fully characterize
$\delta(t)$ provided that the considered time series of sampled
fields are uncorrelated. Two interesting physical interpretations
of $\bar\delta$ can be drawn: first, in a homogeneous turbulent
flow, $\bar\delta=i^2+1$, so that $\bar\delta$ is a global
generalization of the local turbulence intensity; second, for real
flows such as the von K\'arm\'an flow, $\bar\delta$ contains an
additional information regarding how far the instantaneous flow is
from the mean flow. Indeed,
\begin{eqnarray*}
\left(\bar\delta-1\right) \langle \overline{V}^2 \rangle
&=&\overline{\langle (V-\overline{V})^2\rangle}\nonumber\\
&=& \frac{1}{\cal{V}} \overline{\int_{\cal{V}} d^3\vec
x\left(V-\overline{V}\right)^2},
\end{eqnarray*}
where the sum runs over the volume $\cal{V}$ of the flow. The
quantity under the overline is nothing but the square of the mean
distance (using norm 2) between the instantaneous flow and the
mean flow in the functional space. Therefore, the quantity
$\bar\delta-1$ measures how far, on average, the instantaneous
velocity field is from its time average. If $\bar\delta$ is close
to one, one therefore expects the instantaneous flow to strongly
resemble the mean flow. If $\bar\delta$ is much greater than $1$,
the instantaneous field will be more remote from the mean flow.
This interpretation of $\delta$ will be used in appendix
\ref{convergence} to draw a rough study of the convergence of the
von K\'arm\'an flow toward its time average.

\section{Experimental setup and data processing}
\label{secexp}
\subsection{The von K\'arm\'an flow}
\subsubsection{Experimental setup}

In order to illustrate and apply these concepts, we have worked
with a specific axisymmetric turbulent flow: the von K\'arm\'an
flow generated by two counter-rotating impellers in a cylindrical
vessel. The cylinder radius and height are respectively $R=100$ mm
and $240$ mm. We have used two sets of impellers named TM60 and
TM73 \footnote{These impellers have been historically designed for
efficient dynamo action in the von K\'arm\'an Sodium (VKS)
experiment held in CEA-Cadarache. Impellers TM60 were designed,
studied \cite{marie2003} and tested in a first sodium experimental
setup, called VKS1, during years 2000-2002 \cite{bourgoin2002}.
This setup did not succeed in producing dynamo action. Further
optimization process led to TM73 impellers design
\cite{ravelet2005,monchaux07c} and to a new experimental setup
VKS2. In this system, successful dynamo action has been achieved
in 2006 \cite{monchaux2007}.}. These two models are flat disks of
respective diameter $185$ and $150$ mm fitted with radial blades
of height $20$ mm and respective curvature radius $50$ and $92.5$
mm. The inner faces of the
discs are $H=180$ mm apart.\\
Impellers are driven by two independent motors rotating up to
typically $10$~Hz. More details about the experimental setup can
be found in \cite{ravelet2005}. The motor frequencies can be
either set equal to get exact counter-rotating regime, or set to
different values $f_1\ne f_2$. We define two forcing conditions
associated with the concave ({\it resp.} convex) face of the
blades going forward, denoted in the sequel by sense $(-)$ ({\it
resp.} $(+)$ ). We also work with two different vessel geometries,
allowing the optional insertion of an annulus ---thickness $5$ mm,
inner diameter $170$ mm--- in the equatorial plane. Both the
forcing condition and the annulus insertion strongly influence the
level of fluctuation in the flow, thereby allowing to test
experimentally the sensitivity and the relevance of $\delta(t)$,
$\bar\delta$ and $\delta_2$. The working fluid is water.

\subsubsection{Control parameters}
From the two motor frequencies, $f_1$ and $f_2$, we define two
control parameters: a Reynolds number, $Re$, and a rotation
number, $\theta$. For the experiments described in this paper, the Reynolds number,
$$Re=\pi (f_1+f_2) R^2 \nu^{-1},$$ where $\nu$ is the fluid
kinematic viscosity, ranges from $1.25\times 10^5$ to $5\times
10^5$ so that the turbulence can be considered fully developed.
The rotation number,
$$\theta=\frac{f_1-f_2}{f_1+f_2},$$ measures the relative
influence of global rotation over a typical turbulent shear
frequency. Indeed, the exact counter-rotating regime corresponds
to $\theta=0$ and, for a non-zero rotation number, our
experimental system is similar, within lateral boundaries, to an
exact counter-rotating experiment at frequency $f=(f_1+f_2)/2$,
with an overall global rotation at frequency $(f_1-f_2)/2$
\cite{theselouis,ravelet2004}. In our experiments, we vary
$\theta$ from $-1$ to $+1$, exploring a regime of relatively weak
rotation to shear ratio.

\subsubsection{Mean flow topology\label{mft}}
In the exact counter-rotating regime, i.e. for $\theta=0$, the
standard mean flow is divided into two toric recirculation cells
separated by an azimuthal shear layer (cf. Fig. \ref{Map}(b) and
(e)). As $\theta$ is driven away from zero, a change of topology
occurs at a critical value $\theta_c$: the mean flow bifurcates
from the two counter-rotating recirculation cells to a single cell
\cite{ravelet2004}. $\theta_c$ depends on the forcing and the
geometry. For example, in the configuration with TM73 impellers,
rotation sense $(+)$ and the annulus, we measure $\theta_c=0.17\pm
0.01$ through torque measurements. In the case TM60$(-)$, this
turbulent bifurcation becomes highly singular and gives rise to
multistability between the two turbulent flow symmetries,
$O(2)$/$SO(2)$ \cite{ravelet2004}.

\subsection{Measurements and data processing}
Measurements are done with a Stereoscopic Particle Image
Velocimetry (SPIV) system. The SPIV data provide the radial, axial
and azimuthal velocity components on a 95$\times$66 points grid
covering a whole meridian plane of the flow through a time
series of about 5000 regularly sampled values.
The sampling frequency is set between 1 and 4 Hz, corresponding to
one sample record every 1 to 10 impeller rotations. The total
acquisition time is about ten minutes, i.e. one order of magnitude
longer than the characteristic time of the slowest patterns of the
turbulent flow. Fast scales are statistically sampled.\\
The velocity fields are non-dimensionalized using a typical
velocity $V_0=2\pi R(f_1+f_2)/2$ based on the radius of the
cylinder and the rotation frequencies of the impellers. The
resulting velocity fields are windowed so as to fit  to the
boundaries of the flow and remove spurious velocities measured in
the impellers and the boundaries. The resulting fields consist of
58$\times$58 points velocity maps. Two types of filtering are
further applied to clean the data: first, a global filter to get
rid of all velocities larger than $3\times V_0$; then, a local
filter (based on velocities of nearest neighbours) to remove
isolated spurious vectors. Typically, 1~$\%$ of the data are
changed by this processing.

\subsection{Data analysis}
We use two different methods to compute $\delta(t)$ from our PIV
measurements. In the direct method, we compute $\delta$ by
spatially averaging the kinetic energy density of instantaneous
and time averaged flows. Since we measure the full velocity field
in a single meridian plane only, we compute 3D spatial averages of
any quantity $X$ assuming the statistical axisymmetry of the von
K\'arm\'an flow such as:
$$\langle X(r,z) \rangle=\frac{1}{R^2H}\int^R_{-R}|r|dr\int^{H/2}_{-H/2}dz \,X(r,z).$$
Since velocity fields are discrete in space and time, spatial
integration is done with a classical numerical trapezoid summation
method whereas time integration is performed through simple
summation. The parameter $\delta$ can also be obtained by a
spectral method. For this, we compute the spatial Power Spectral
Density (PSD) of the discrete velocity fields as:
$$\widetilde{E}(k_r,k_z,t)=\widetilde{W}\,\widetilde{W}^*,$$ where
$\widetilde{W}=\widetilde{W}(k_r,k_z,t)$ is the two dimensional
spatial Fourier transform of $W(r,z,t)=\sqrt{|r|}\,V(r,z,t)$ and
$\widetilde{W}^*$ its complex conjugate. Then, we compute $\delta$
using Parseval's identity:
$$\int\int |r|drdz\,V^2=\int\int dk_rdk_z \widetilde{W}\widetilde{W}^*,$$
and get:
\begin{equation}\label{deltadekinteg}
\delta(t)= \frac{\int\int dk_rdk_z \widetilde{E}(k,t)}{\int\int
dk_rdk_z \,\widetilde{E}_0(k)},
\end{equation} where $\widetilde{E}_0$ is the PSD of the
time averaged flow
$$\widetilde{E}_0(k_r,k_z,t)=\widetilde{\overline{W}}\,\widetilde{\overline{W}}^*.$$

\section{Results and applications}
\label{appli}

\subsection{Basic properties of $\delta$}

\subsubsection{Statistical properties}
\begin{figure}[h]
\centerline{\includegraphics[width=0.9\columnwidth]{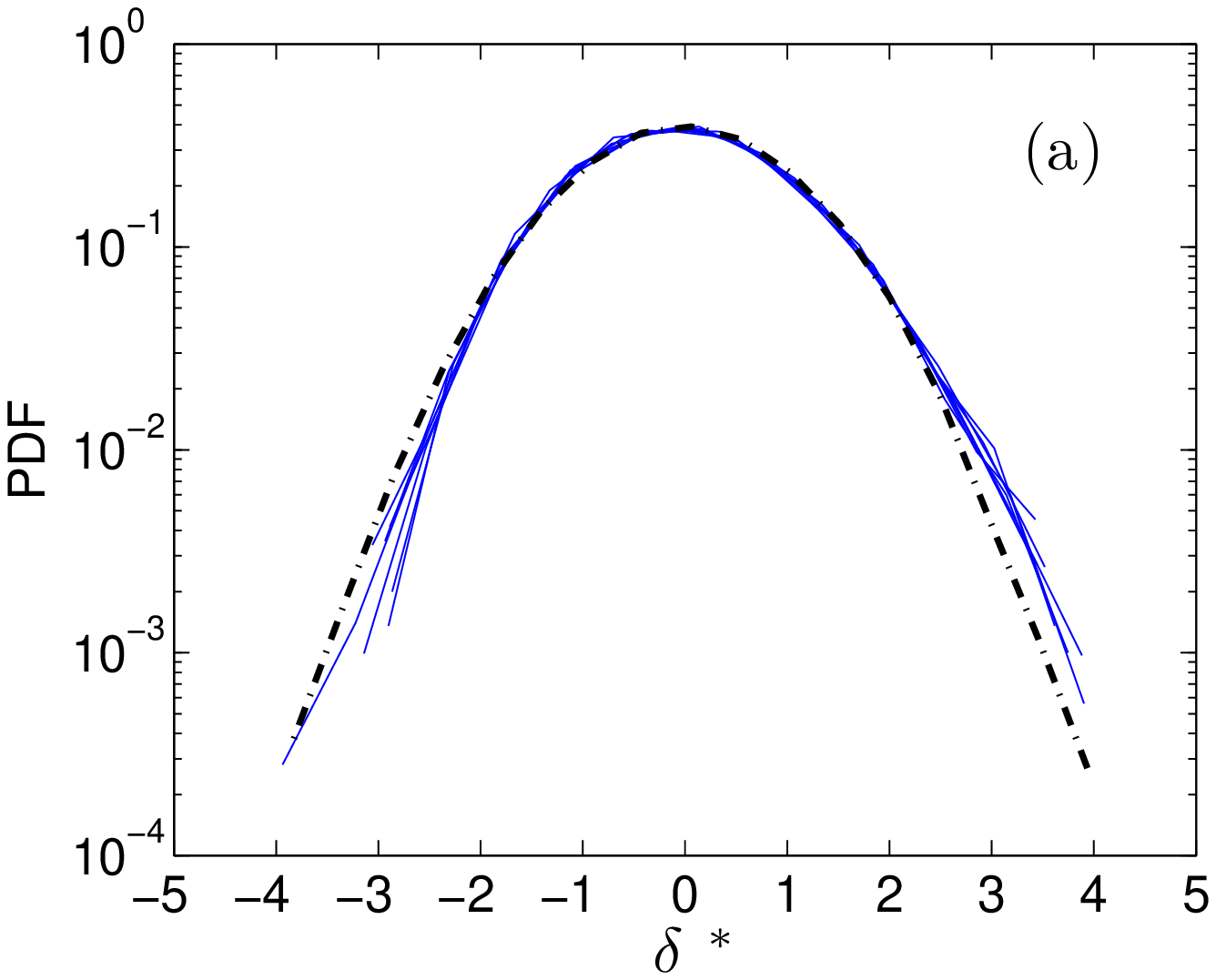}}
\centerline{\includegraphics[width=0.9\columnwidth]{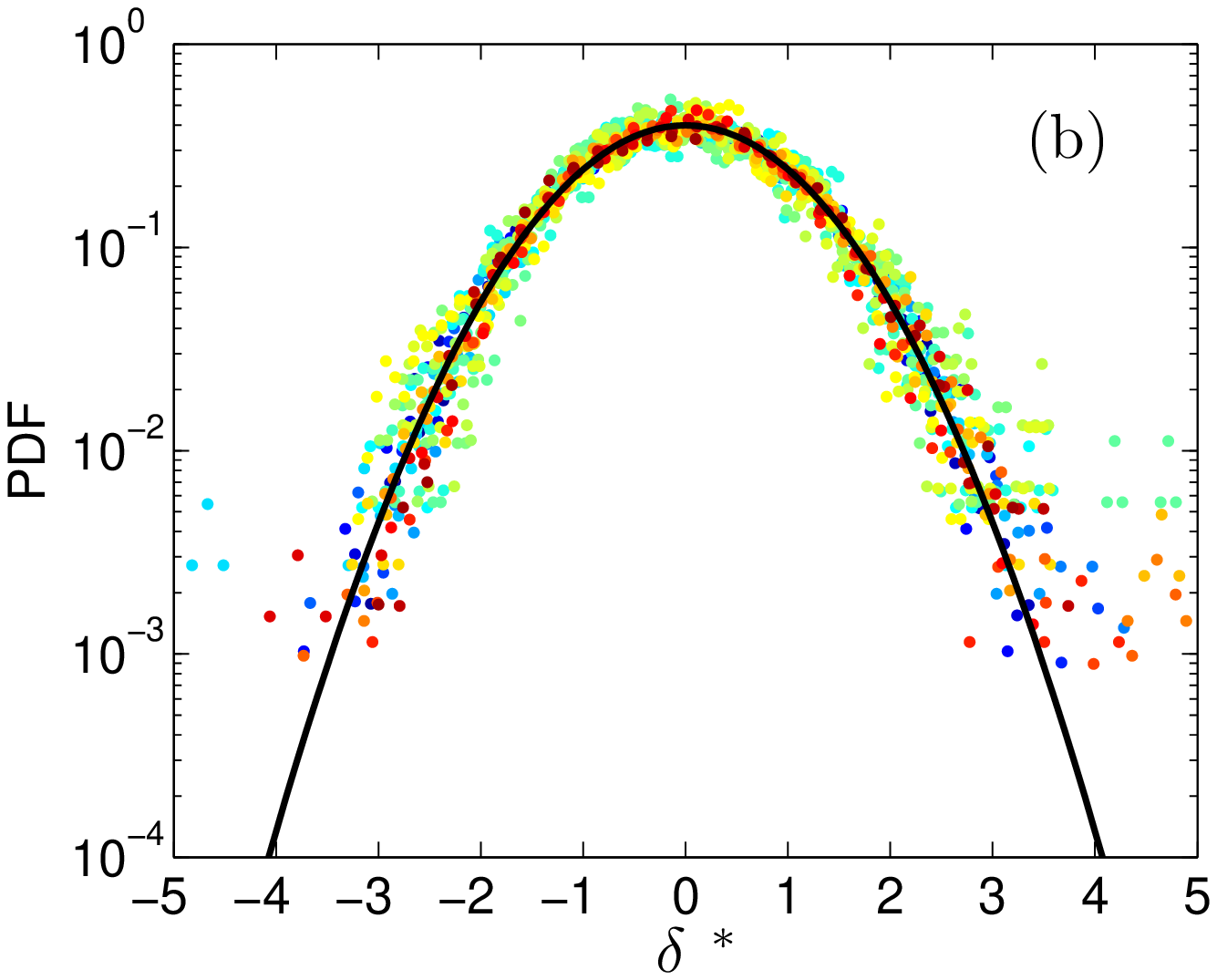}}
\caption{Centered and reduced Probability Density Function of
$\delta(t)$, (a) for the TM73$(-)$ configuration with annulus at
seven different Reynolds numbers ranging from $1.25 \times 10^5$
to $5\times 10^5$, and (b) for the TM73$(+)$ configuration without
annulus for $16$ experiments performed with different values of
rotation number $\theta=(f_1-f_2)/(f_1+f_2)$ ranging from $0$ to
$1$. $\delta^*=(\delta-\bar\delta)/\delta_2$ is the centered
reduced value of $\delta$. The dash-dotted lines are Gaussian
functions of zero mean and unit variance.} \label{pdfconv}
\end{figure}

As discussed in Sec. \ref{theor}, $\delta(t)$ widely fluctuates in
time. However, its statistical properties in the exact
counter-rotating case are rather simple since its probability
density function (PDF) is nearly Gaussian as illustrated in Fig.
\ref{pdfconv}(a). Actually, this property holds at different
Reynolds numbers (cf. Fig. \ref{pdfconv}(a)) and for all forcing
and geometry conditions studied in the present paper (cf. Fig.
\ref{pdfconv}(b)). A striking feature is that global rotation does
not change the statistical properties of the flow since the PDFs
are Gaussian at any $\theta$. Consequently, we can describe and
fully characterize the total energy temporal distribution using
only the two scalar parameters $\bar\delta$ and $\delta_2$.

\subsubsection{Dependence on the Reynolds number}

\begin{figure}
\psfrag{d}[c][][1.2]{$\bar\delta$}
\includegraphics[width=\columnwidth]{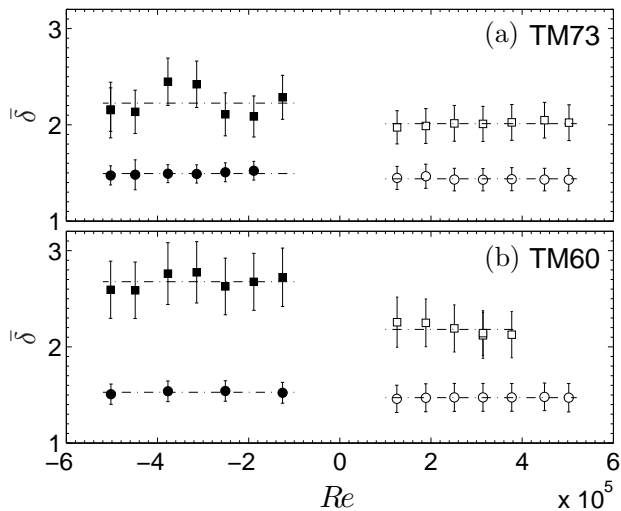}
\caption{$\bar\delta$ as a function of Reynolds number $Re$.
Negative $Re$ correspond to rotation sense $(-)$. (a) and (b)
correspond respectively to TM73 and TM60 impellers, with
($\circ$), and without ($\scriptstyle \square$) annulus.
Dash-dotted lines corresponds to the mean value of $\bar\delta$
for each configuration and $\delta_2$ is represented using error
bars.} \label{deltaTM}
\end{figure}

For a given forcing, we observe a very weak dependence of
$\bar{\delta}$ with $Re$ in the studied range as one can see in
Fig. \ref{deltaTM}. The largest dispersions, of the order of
$8\%$, are observed in the negative rotation sense for both TM60
and TM73 impellers without annulus. For any other forcing
condition, the dispersion is less than $5\%$. Similar conclusions
hold for $\delta_2$ represented in Fig.~\ref{deltaTM} using error
bars. This behavior is not surprising since in our water
experiments the Reynolds number is so large that we are in the
fully developed turbulent regime where statistical properties have
already been shown not to depend on the Reynolds number
\cite{monchaux2006,ravelet2008,monchaux07c}. On the contrary, we
expect a different behavior when lowering the Reynolds number
\cite{monchaux2006}, since, in the limit of very low $Re$, the
instantaneous flow is laminar and identical to the mean flow:
$\delta(t)=\bar\delta=1$. The study of the transition between
laminar and fully turbulent regime requires the use of
glycerol-water mixing and is beyond the scope of the present
paper. Note however that this transition has already been studied
in the von K\'arm\'an experiment ---TM60$(-)$ without annulus---
by Ravelet et al. \cite{ravelet2008} using a one point measurement
Laser Doppler Velocimetry system. It has been shown that the local
value of the variance of the azimuthal velocity,
$\overline{v_\theta^2}-\overline{v}_\theta^2$, behaves as
$(Re-Re_c)^{1/2}$ and saturates above $Re_t=3300$, where
$Re_c=330$ corresponds to the transition from steady to
oscillatory laminar flow and $Re_t$ to the onset of the fully
turbulent inertial regime. A similar behaviour is encountered in
Direct Numerical Simulations of the Taylor-Green flow
\cite{LavalDyn} where an increase of $\bar\delta$ from 1 at low
$Re$ to a saturation value of 3 for $Re$ above $10^3$ has been
observed.

\subsection{Influence of forcing and annulus \label{forcing&annulus}}

It is difficult to estimate turbulence intensity by simply looking
at an instantaneous velocity field (cf. Figs. \ref{Map}(c) and
(f)). To achieve this, the local turbulence intensity $i$ maps
(cf. Figs. \ref{Map}(a) and (d)) constitutes a good qualitative
tool to estimate the overall turbulence intensity and structure of
the considered flow. However, a quantity like $\delta$ is needed
to provide a global quantitative tool.

\subsubsection{Effect of impeller and vessel geometry}

As reported in Table \ref{tabDelta}, the values of $\bar\delta$
and $\delta_2$ allow to quantify the differences in the turbulence
fluctuations level as a function of the forcing configurations
(TM60 and TM73, with or without annulus).
\begin{table}[htdp]
\begin{center}
\begin{tabular}{|*{9}{c|}}
  \hline
    & \multicolumn{4}{|c|}{$\bar\delta$} &  \multicolumn{4}{|c|}{$\delta_2$} \\
   \hline
   Impellers & \multicolumn{2}{|c|}{TM60} &  \multicolumn{2}{|c|}{TM73} & \multicolumn{2}{|c|}{TM60} &  \multicolumn{2}{|c|}{TM73}  \\
  \hline
   Sense & $(-)$ & $(+)$  & $(-)$  & $(+)$ & $(-)$ & $(+)$  & $(-)$  & $(+)$  \\
   \hline
   Without annulus & 2.64 & 2.18 & 2.22 & 2.02 &0.30&0.24&0.24&0.18\\
   \hline
   With annulus & 1.52 & 1.47 & 1.50& 1.48 &0.11&0.15&0.11&0.12\\
   \hline
\end{tabular}
\end{center}
\caption{Values of $\bar\delta$ and $\delta_2$ for various
configurations.} \label{tabDelta}
\end{table}
The major result of Table \ref{tabDelta} is the effect of the
annulus which systematically reduces $\bar \delta$ and $\delta_2$,
meaning that the instantaneous flow is much closer to the mean
flow. A great part of this drop probably reflects the reduction of
the slow fluctuations of the shear layer already known from time
spectral analysis to be responsible for a major part of the energy
fluctuations \cite{ravelet2008}. From a spatial point of view,
this reduction is the consequence of the expected locking of the
shear layer in the annulus plane which appears in direct
visualizations of the flow. Additionally, we note that the annulus
tends to collapse all values of $\bar\delta$ to $1.5\pm0.03$. In
contrast, a dispersion of about thirty percents remains on the
variance $\delta_2$.

\subsubsection{Qualitative interpretation}
\label{Vortex}

To understand the measures presented in the last paragraph, we
report visual observations of the flow seeded with bubbles. Such
visualizations allow to unveil spatial structures of turbulence
and especially the largest patterns which exhibit the slowest time
dynamics. Let us describe our observations in the TM73 impellers
configuration.

\begin{figure}
\includegraphics[width=0.9\columnwidth]{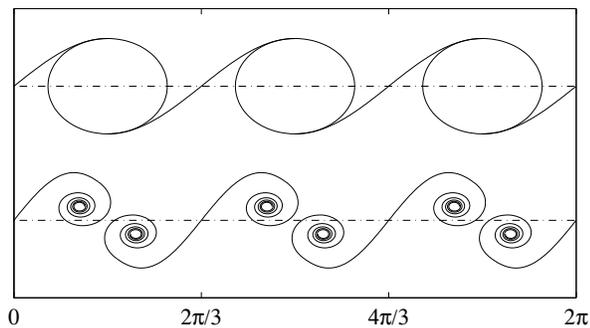}
\caption{Sketch of the shear-layer patterns, on the developed
cylinder $(\theta, z)$, from visual observations. Without annulus
(top sketch), the shear produces a triplet of large corotating
vortices, corresponding to a $m=3$ mode. In the presence of the
equatorial annulus (bottom sketch), the three vortex cores split
into corotating pairs: there is a strong energy transfer from the
fundamental $m=3$ mode to its first $m=6$ harmonics.}
\label{vortexfig}
\end{figure}

Without annulus, the main structure of the free shear layer at
very high Reynolds number consists of three big fluctuating
vortices, i.e. a $m=3$ azimuthal wave number mode (cf. Fig.
\ref{vortexfig}). The size of these vortices is almost the full
free height of the vessel for sense $(-)$ and just a little bit
smaller for sense $(+)$. These vortices fluctuate in azimuthal
position as well as in amplitude or apparent size. Nucleation and
merging frequently occur.

In the presence of an annulus, each vortex splits into a pair of
smaller co-rotating vortices attached to the leading edges of the
annulus (see the schematic view in Fig. \ref{vortexfig}). The
$m=6$ harmonics appears and pumps an important amount of energy to
the fundamental $m=3$ mode. The strong reduction in $\bar\delta$
and $\delta_2$ is probably the trace of this scale change in the
shear layer structure. Furthermore, we observe the temporal
dynamics of the shear layer vortices which is different for both
senses of rotation. For sense $(+)$, the vortices keep high
mobility and fluctuation levels. On the contrary, for sense $(-)$,
the vortices are almost steady: merging, nucleation and large
azimuthal excursions are inhibited. The only remaining dynamics is
an azimuthal vibration of their center at a frequency of a few
Hertz. This qualitative observation is corroborated by the
behavior of $\delta_2$: even if $\bar\delta\sim 1.5$ in both
cases, $\delta_2$ is lower in the $(-)$ senses, where the vortex
pairs are more stable.

\subsubsection{Scale by scale characterization}

\begin{figure}[htb]
\psfrag{d}[c][][1.1]{$\overline{\widetilde{\delta}}(k)$}
\psfrag{ka}[c][][1.1]{$k$ ($D^{-1}$)}
\begin{center}
\includegraphics[width=\columnwidth ]{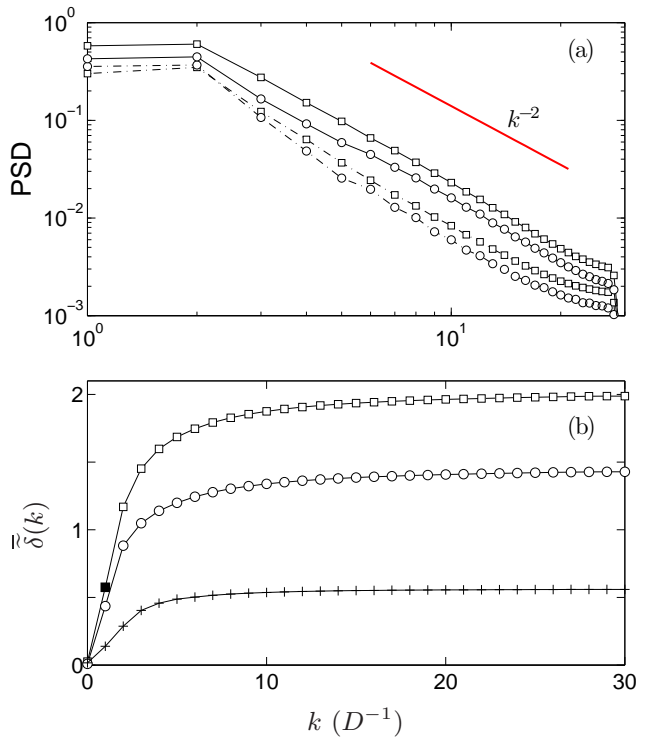}
\caption{(a) Two-dimensional time-averaged Power Spectral Density
of instantaneous flows (solid lines) and Power Spectral Density of
the corresponding time-averaged flows (dash-dotted lines) for two
experiments performed with setup TM73$(+)$, without the annulus
($\scriptstyle \square$) and with the annulus ($\circ$). The PSD
are normalized by the total kinetic energy of the mean flow. (b)
$\overline{\widetilde{\delta}}(k)$ as function of $k$ for the same
setup without ($\scriptstyle \square$) and with annulus ($\circ$).
The specific data point $\overline{\widetilde{\delta}}(1/D)$ is
represented with a black symbol ($\scriptstyle \blacksquare$). The
difference $\Delta\overline{\widetilde{\delta}}(k)$ between the
two curves is also displayed (plusses). At high $k$, it is of the
order of $\overline{\widetilde{\delta}}(1/D)$ without annulus (see
text for details). Wavenumbers are normalized by the PIV window
size $D=1.93 R$.} \label{spec}
\end{center}
\end{figure}

By construction, $\bar\delta$ accounts for the total fluctuating
kinetic energy of the turbulent flow. The use of PSD as an
intermediate step to compute $\bar\delta$ (cf. Eq.
(\ref{deltadekinteg})) can provide useful information about the
distribution of the energy over the various length scales of the
flow as illustrated in Fig.~\ref{spec}. Indeed, in
Fig.~\ref{spec}(a), we compare the time-averaged PSD of the
instantaneous velocity fields, $\overline{\widetilde{E}}(k)$, to
the PSD of the corresponding time averaged field,
$\widetilde{E}_0(k)$, for the two TM73$(+)$ setups, with and
without annulus.

In the spirit of the definitions of turbulent intensity $i$ and of
the quantity $\delta$, and because we do access the full spatial
spectra, we choose to normalize the PSD by the total kinetic
energy of the mean flow. Therefore, the integrals of the displayed
curves for the two time-averaged flows are equal to $1$. The wave
number $k$ is given by: $k^2=k_r^2+k_z^2$.

In the time-averaged PSD of the instantaneous flow, the major part
of the fluctuating kinetic energy is concentrated close to the
forcing scale, where the energy is injected, like in any
homogeneous and isotropic turbulent flow. On the contrary, at
higher $k$, the slopes of the different PSD are close to $-2$ and
larger than the classical $-5/3$ exponent observed for homogeneous
isotropic turbulence.

At first sight, the time-averaging removes kinetic energy at all
spatial scales. A closer look indicates different  relative
reductions depending on the presence or the absence of the
annulus. At high $k$, the ratio of the PSD of instantaneous to
time-averaged flows is of the same order of magnitude for both
configuration. However, at low $k$, this ratio is $2$ (resp.
$1.2$) without (resp. with) annulus: there are five time less
fluctuating low-spatial-frequencies to be averaged when the
annulus is inserted meaning that the instantaneous flow is closer
to the time-averaged flow.

To quantify more precisely the weight of each scale in
$\delta(t)$, we introduce $\widetilde{\delta}(k,t)$ defined as:
$$\widetilde{\delta}(k,t)=\frac{\int^k_0dk'\,\widetilde{E}(k',t)}{\int^\infty_0dk'\,\widetilde{E}_0(k')},$$
so that $\widetilde{\delta}(k,t){\rightarrow}\delta(t)$ when $k
\rightarrow \infty$. In Fig.~\ref{spec}(b),
$\overline{\widetilde{\delta}}(k)$ is plotted for TM73$(+)$ flows
with and without annulus. These curves represent the integral from
$0$ to $k$ of the above normalized PSD and they converge towards
$\bar\delta$ at large $k$.

First of all, we see that the contribution of the 5 first wave
numbers, i.e. of spatial modes larger than $D/5 \simeq R/10$, is
about eighty percents of the total value of $\bar\delta$: again,
large scales dominate the fluctuations.

Some quantitative analysis can be carried on. We note that the
fluctuation level in the first $k=1/D$ mode of the
$\overline{\widetilde{\delta}}(k)$ curve without annulus and the
difference $\Delta\bar{\tilde{\delta}}$ between the two flows
---with and without annulus--- are of the same order of magnitude.
This is visible in Fig.~\ref{spec}(b), where the
$\overline{\widetilde{\delta}}(1/D)$ data point has been
emphasized, and has been verified for the three other couples of
flows. As described in the previous section, the effect of the
annulus can be seen as a spatial filtering of the largest patterns
of the flow. The difference between the two curves can be
described by a simple empirical protocol. The amount of
$\bar\delta$ in excess, i.e. $\overline{\widetilde{\delta}}(1/D)
\simeq \Delta\bar{\tilde{\delta}}$, is subtracted from the curve
$\overline{\widetilde{\delta}}(k)$ without annulus in a geometric
progression over $k$, e.g.
$\overline{\widetilde{\delta}}(1/D)/2^k$ for each wavenumber $k$
starting from $D^{-1}$ or $2 D^{-1}$ depending on the
configuration. The result is a curve really close to
$\overline{\widetilde{\delta}}(k)$ with annulus. The effect of the
annulus may thus be seen as a high-pass filtering, which confirms
the sketch presented in Fig. \ref{vortexfig}. Finally, the
quantity $\widetilde{\delta}(k,t)$ appears as an efficient tool to
track the spectral changes in turbulent scales.

\subsection{Properties of $\delta$ when $f_1\neq f_2$}

As already mentioned in Sec. \ref{mft}, depending on the value of
the rotation number $\theta=(f_1-f_2)/(f_1+f_2)$, the mean von
K\'arm\'an flow exhibits two different topologies. Indeed, for
$|\theta|<\theta_c$, the mean flow is composed of two toric
recirculation cells separated by an azimuthal shear layer, as for
$|\theta|>\theta_c$, it is composed of a single recirculation cell
\cite{ravelet2004}. We have performed a set of experiments for the
specific forcing condition TM73$(+)$ with and without annulus,
varying $\theta$ from $-1$ to $1$. In this geometry, $\theta_c =
0.17 \pm 0.01$ with annulus and $\theta_c \simeq 0.095\pm 0.005$
without annulus. The aim of these experiments is to study the
influence of the global rotation and of the transition between the
two flow topologies on turbulence intensity.

\begin{figure*}
\psfrag{p}{shear layer position}
\psfrag{a}[c][][1.1]{(a)}\psfrag{b}[c][][1.1]{(b)}\psfrag{c}[c][][1.1]{(c)}\psfrag{d}[c][][1.1]{(d)}\psfrag{e}[c][][1.1]{(e)}\psfrag{f}[c][][1.1]{(f)}
\psfrag{d1}[c][][1.1]{$\bar\delta$}\psfrag{d2}[c][][1.1]{$\delta_2$}
\begin{tabular}{cc}
    without annulus & with annulus\\
    {\includegraphics[width=\columnwidth]{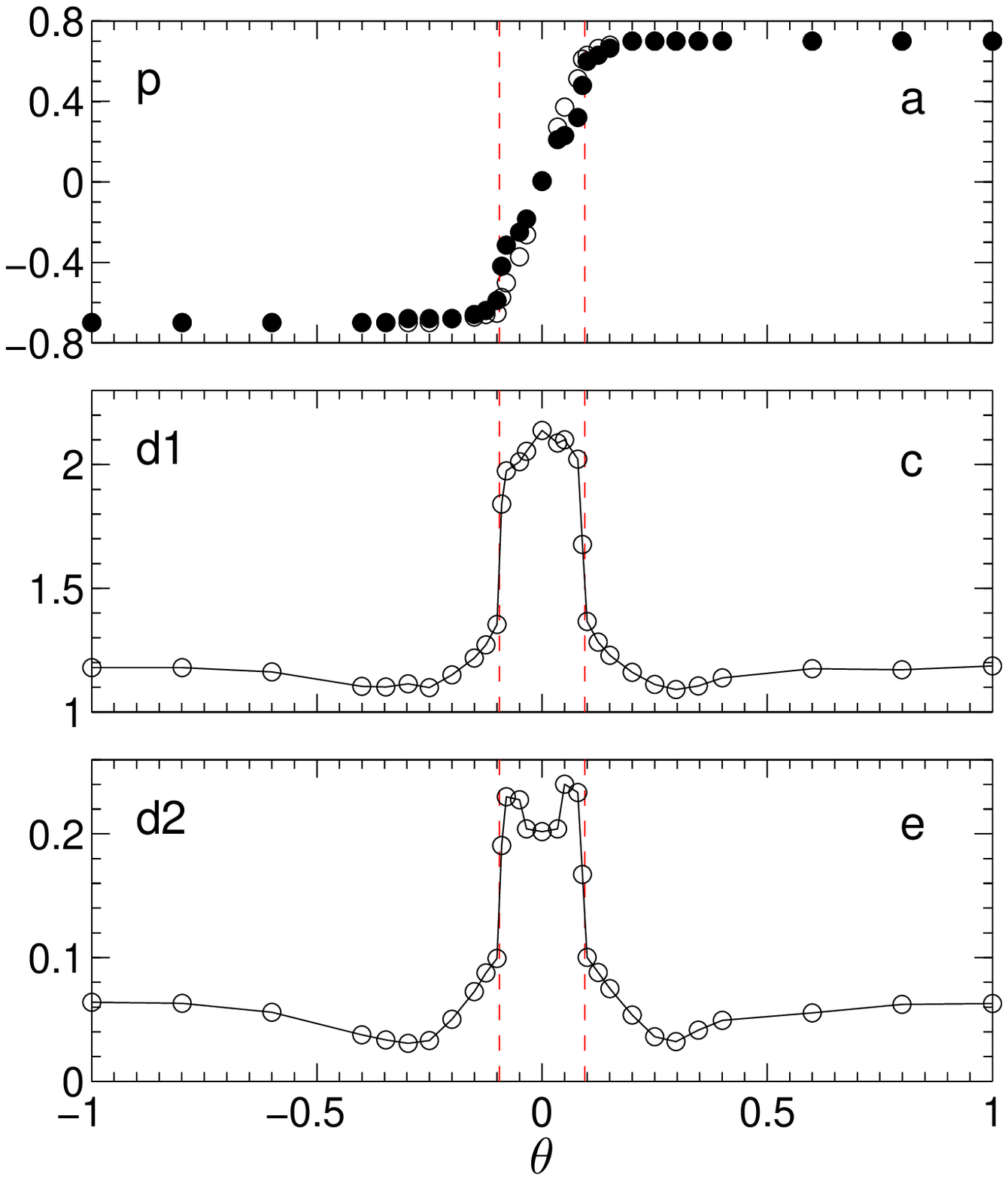}} &
    {\includegraphics[width=\columnwidth]{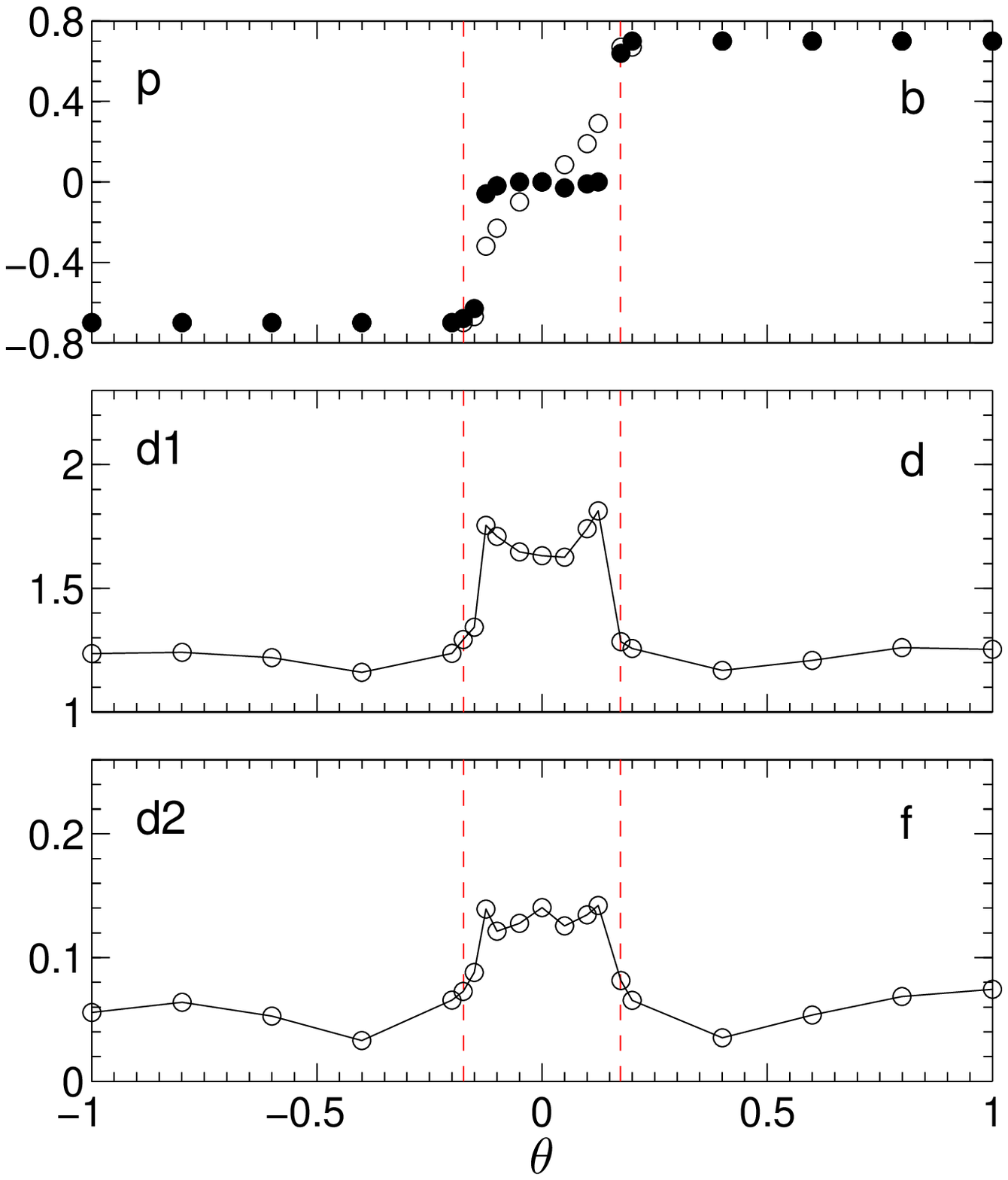}} \\
\end{tabular}
\caption{\label{theta}(a) and (b), vertical position of the shear
layer at $r=0$ on the rotation axis ($\circ$) and at position
$r=0.7$ ($\bullet$), (c) and (d) $\bar\delta$, and (e) and (f)
$\delta_2$, as a function of the rotation number $\theta$. Left
and right columns correspond respectively to setup TM73$(+)$
without and with annulus. In (a) and (b), the ($\circ$) markers
correspond to the stagnation point vertical position. In (c-f),
$\bar\delta$ and $\delta_2$ are computed over the whole flow.}
\end{figure*}

A way to quantify precisely the turbulent bifurcation of the mean
flow occurring at $\theta_c$ is to study the position of the
azimuthal shear layer that separates the recirculation cells when
$|\theta|<\theta_c$. A good quantity that allows to localize this
position is the zero iso-surface of the stream function
$\psi(r,z)$ of the mean flow defined through
$(\overline{u_r},0,\overline{u_z})= \nabla\times
(r^{-1}\overline{\psi}\,{\bf e}_\theta)$ in cylindrical
coordinates. Thus, in Figs. \ref{theta}(a) and (b), we have
plotted two measures of the shear layer vertical position at
$r=0$, i.e. the stagnation point, and at $r=0.7$ on the side. The
discrepancies between these two sets of data renders the fact the
shear layer is in general a curved surface expect for the
$\theta=0$ configuration. Actually, its vertical position at large
$r$ is always closer to the $z=0$ equatorial plane of the vessel
than at $r=0$. As $|\theta|$ increases from zero, the shear layer
is attracted by the slowest impeller. Without annulus (cf. Fig.
\ref{theta}(a)), it results in a global drift of the shear layer
\cite{cadot07} accompanied by the appearance of a moderate
curvature. With annulus (cf. Fig. \ref{theta}(b)), the outer edge
of the shear layer remains pinned on the annulus whereas its
center ---the stagnation point--- drifts, increasing the layer
curvature up to a high level just below $\theta_c$: the annulus
strongly stabilizes the shear layer near the equatorial plane and
maintains it to higher rotation numbers.

At $\theta=\theta_c$, we observe a discontinuity of the shear
layer vertical position which corresponds to the turbulent
bifurcation. The discontinuous bifurcation does not present, in
the two studied cases, any hysteresis in $\theta$. However, close
to $\theta_c$, we can observe transitions between two metastable
one cell and two cells states which follow a slow dynamics
\footnote{Careful analysis of the critical regimes by torque
measurements, to be reported elsewhere, has been performed with
the annulus. In a very narrow range $0.171 \lesssim \theta\
\lesssim 0.179$, we observe very slow dynamical regimes where the
topology changes back and forth along time over hours. This
explains why, very close to $\theta_c$, some of the present
measurements ---acquired for only 10 minutes--- may appear of the
wrong topology in Fig. \ref{theta}.}.

In this section, we first analyze how the parameters $\bar\delta$
and $\delta_2$ behave with the rotation number. Thereafter, using
these tools computed over the two symmetric half of the flow, we
analyze how they can also provide a proper characterization of the
symmetry breaking.

\subsubsection{Variation of $\delta$ with the rotation number}

The parameter $\delta$ can be used to study the changes in
turbulence intensity related to the turbulent bifurcation
undertaken by von K\'arm\'an flows at critical $\theta_c$. Figs.
\ref{theta} (c-f) show the variations of $\bar\delta$ and
$\delta_2$, calculated over the whole flow as a function $\theta$.

First of all, for $|\theta|<\theta_c$, $\bar\delta$ and $\delta_2$
are relatively high ($\bar\delta \simeq 2$ and $\delta_2 \simeq
0.2$). This reflects the presence of the highly fluctuating shear
layer in the flow bulk. On the contrary, for $|\theta|>\theta_c$,
i.e. for a large rotation-to-shear rate, $\bar\delta$ and
$\delta_2$ are quite small ($\bar\delta \simeq 1.2$ and $\delta_2
\simeq 0.05$) and are almost independent of $\theta$ \footnote{If
regions of high level of turbulence due to shear still exist, they
should be confined inside the blades of the slow rotating
impeller, where measurements are impossible to carry on.}. This
means that the level of global rotation has no significant
influence on the turbulence intensity level unless the mean flow
topology changes. Actually, $\bar\delta$ and $\delta_2$ carefully
trace back the flow topology changes induced through the
bifurcation at $\pm \theta_c$. They might even be used as order
parameters of such turbulent bifurcation and they provide a
reliable measurement of threshold $\theta_c$.

\subsubsection{Effect of the annulus}

We have seen how the annulus stabilizes the shear layer and moves
the bifurcation threshold. It has also a strong effect on the
turbulence level. In Figs. \ref{theta} (c-f), we observe that for
$|\theta|<\theta_c$, when the flow is composed of two
counter-rotating toroidal cells, the turbulence level is much
larger without than with the annulus: the results for
zero-rotation-number (cf. Sec. \ref{forcing&annulus}) extend to
$-\theta_c < \theta < \theta_c$.

We have already mentioned that this strong reduction in $\bar
\delta$ and $\delta_2$ with the annulus is due to the
stabilization, especially at large scales, of the shear layer that
is trapped by the annulus. Now, picturing that the flow has a
given energy gap to overcome in order to bifurcate from two cells
to one cell, the lower level of fluctuations in the case with
annulus explains why it is necessary to explore larger $\theta$
before turbulent bifurcation occurs, and then, why $\theta_c$ is
larger with the annulus. Actually, the annulus is postponing the
bifurcation in $\theta$.

For $|\theta|>\theta_c$, the low turbulence intensity, almost
unchanged at first order, is slightly larger in the presence of
the annulus: the one-cell flow is almost insensitive to the
annulus presence and the slight increase is probably due to the
vertical step flow over the annulus.

\subsubsection{Quantifying the symmetry breaking}

\begin{figure*}
\psfrag{a}[c][][1.1]{(a)}\psfrag{b}[c][][1.1]{(b)}\psfrag{c}[c][][1.1]{(c)}\psfrag{d}[c][][1.1]{(d)}
\psfrag{d1}[c][][1.1]{$\bar\delta$}\psfrag{d2}[c][][1.1]{$\delta_2$}
\begin{tabular}{cc}
    without annulus & with annulus\\
    {\includegraphics[width=\columnwidth]{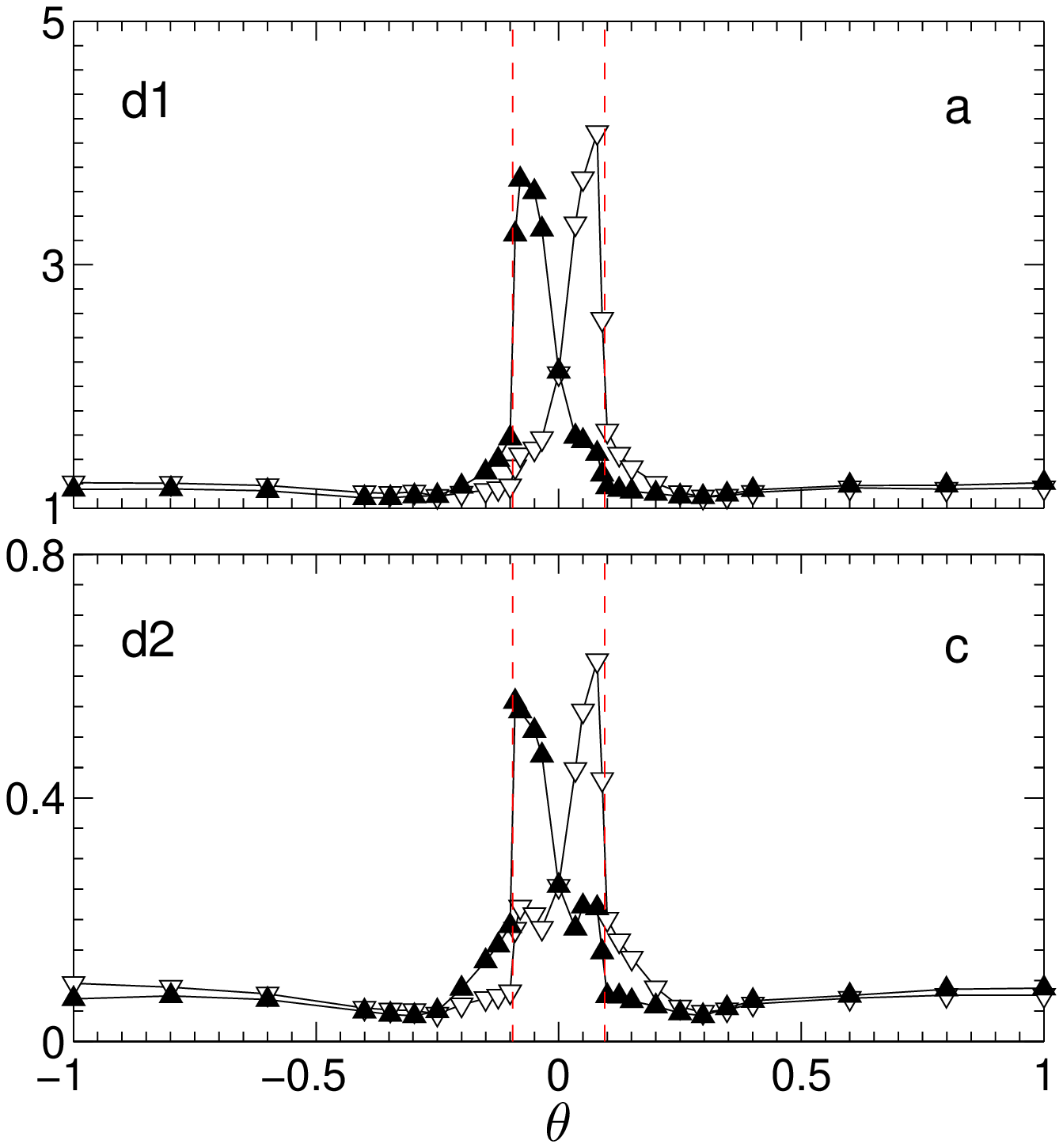}} &
    {\includegraphics[width=\columnwidth]{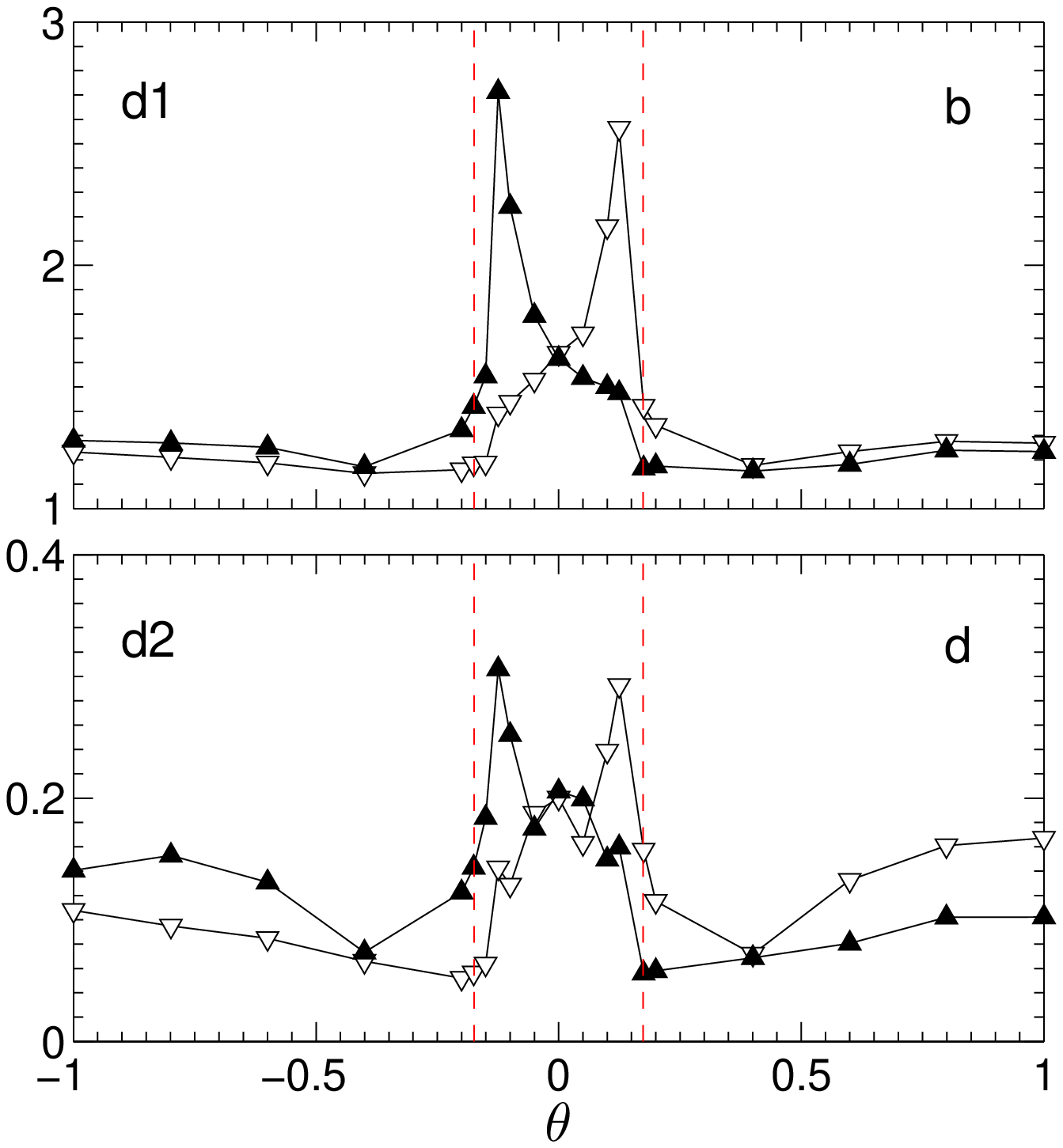}}
\end{tabular}
\caption{\label{thetabis}(a) and (b) $\bar\delta$, and (c) and (d)
$\delta_2$, as a function of the rotation number $\theta$. Left
and right columns correspond respectively to setup TM73$(+)$
without and with annulus. ($\blacktriangle$) correspond to
$\delta$ computed over the upper half part of the flow as
($\triangledown$) to the lower half part.}
\end{figure*}

For further exploration of the turbulent bifurcation and related
symmetry breaking, we now calculate $\bar \delta$ and $\delta_2$
over half vessels, i.e. on each side of the annulus equatorial
plane (cf. Fig. \ref{thetabis}).

Starting from the exact counter-rotating configuration, as
$|\theta|$ is increased from zero, $\bar \delta$ and $\delta_2$
are becoming larger for the half part of the flow corresponding to
the slowest impeller and are decreasing for the other half part.
This is directly connected with the position of the shear layer
which gets closer and closer to the slowest impeller when $\theta$
increases (cf. Figs. \ref{theta}(a) and (b)). This displacement
goes on until the level of fluctuations is strong enough to
activate the transition of the flow from two to one cell at
$\theta=\theta_c$.

Indeed, we clearly see in Figs. \ref{thetabis} the peaks in $\bar
\delta$ and $\delta_2$ that correspond exactly to the bifurcation
point $\theta=\theta_c$. With the annulus, where the bifurcation
is much sharper, one could detect a critical divergence for
$(\bar\delta-1)$ (cf. Fig. \ref{thetabis}(b)).

\section{Conclusive discussion}

We have introduced a new global quantity $\delta$ that
characterizes turbulent fluctuations in inhomogeneous anisotropic
flows. This time-dependent quantity is based on spatial averages
of global velocity fields rather than classical temporal averages
of local velocities. We have shown that, generally, properties of
$\delta(t)$ seems to be fully provided thanks to only two
parameters, its time average $\bar \delta$ and variance
$\delta_2$. These parameters generalize to inhomogeneous
anisotropic flows the classical notion of turbulence intensity,
based on local, single point measurements.

Properties of $\bar \delta$ and $\delta_2$ have been
experimentally studied in the typical case of the von K\'arm\'an
flow for different forcing and geometries. We have shown that, in
the fully turbulent regime, they are Reynolds independent, like
any classical inertial range quantity. However, $\bar \delta$ and
$\delta_2$ depend on forcing and geometry and faithfully reflect
major changes in the flow topology. They can therefore be used as
a new tool for comparison of different turbulent flow. In the
present paper, we provided an example in which $\delta$ is used to
characterize the turbulent bifurcation in the von K\'arm\'an flow
induced through differential rotation of the two impellers.
Finally, $\bar \delta$ and $\delta_2$ are used to characterize the
scale by scale energy budget as a function of the forcing mode as
well as the transition between the two flow topologies.

Another interesting application would be the study of the
generation of magnetic field by a turbulent flow, the so-called
dynamo instability. This problem has attracted recently a lot of
experimental attention
\cite{gailitis01,stieglitz01a,monchaux2007}. In the case where the
corresponding turbulent flow has a non-zero mean value, the dynamo
instability may be seen as a classical instability problem,
governed by the mean flow and the fluctuations
\cite{leprovost2005,LavalDyn,petrelis06b}. A natural question in
this case is therefore how to quantify the relative level of
fluctuations, and the deviations from the mean flow they induce,
so as to implement efficient control strategies to decrease or
increase their influence. Since our global parameter quantify the
difference between the instantaneous and the mean flow, they are
natural candidate to discriminate between different dynamos. This
has been recently illustrated in numerical simulations of
Taylor-Green flow \cite{dubrulle07}. No equivalent quantitative
analysis has been performed for experimental dynamos. However,
among these successful experimental dynamos, we can suspect that
the Karlsruhe \cite{stieglitz01a} and Riga \cite{gailitis01} flows
are characterized by values of $\delta$ lower than in von
K\'arm\'an flows. In the future, we plan to use these parameters
for the analysis of recent results of the von K\'arm\'an Sodium
(VKS) experiment.

\appendix

\section{Convergence towards the mean flow}\label{convergence}

\begin{figure}[h]
\psfrag{d}[c][][1.1]{$\bar{\delta}(N)-1$}
\includegraphics[width=.90\columnwidth]{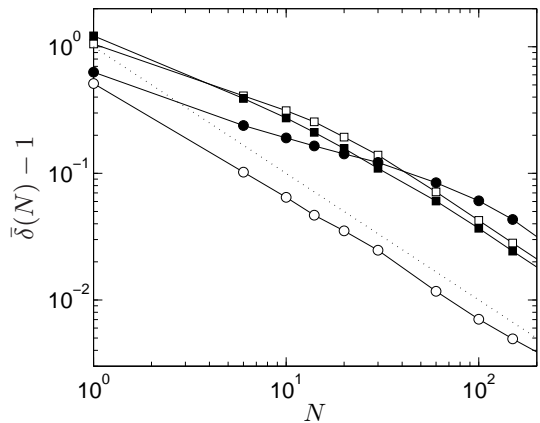}
\caption{$\bar\delta-1$ computed with velocity fields averaged
over $N$ instantaneous fields for various setups : TM73$(-)$,
without the annulus ($\scriptstyle \blacksquare$) and with the
annulus ($\bullet$) and TM73$(+)$, without annulus ($\scriptstyle
\square$) and with annulus ($\circ$). A guide to the eye for the
$N^{-1}$ dependency is plotted with a dotted line.}
\label{deltaconv}
\end{figure}
As we have seen in Sec. \ref{theor}, the parameter $\bar\delta$
can be seen as the average square distance between the
instantaneous and the time averaged velocity fields. Therefore, we
can use a slight modification of $\bar\delta$ to study the
convergence towards the mean flow through statistical averaging.
For this, we define $V(N,t)$ the velocity field averaged around
time $t$ over $N$ instantaneous fields. $N<N_{max}$ where
$N_{max}=5000$ is the total number of instantaneous fields. From a
practical point of view, $V(N_{max})=\overline{V(t)}$. Then, we
define:
\begin{equation}
\bar\delta(N)=\frac{\overline{\langle V^2(N,t) \rangle}}{\langle
\overline{V}^2 \rangle}. \label{convergencedelta}
\end{equation}
With this definition, $\bar\delta(1)=\bar\delta$ and
$\bar\delta(N_{max})=1$. Moreover, $\bar\delta(N)-1$ measures the
average square distance between the partially averaged field
$V(N)$ and the mean flow $\overline{V}$, so that its variations
with $N$ can be used to study the convergence towards the mean
flow. We can see in Fig.~\ref{deltaconv} that this square
distance, $\bar\delta(N)-1$, decreases as $N^{-1}$, at least at
large $N$, what is typical of an uncorrelated fluctuating
quantity. However, for the negative-rotation-sense-with-annulus
case, the $N^{-1}$ dependency is observed only at $N$ larger than
$100$. Actually, in that particular setup, we have observed, by
means of bubble seeding (cf. end of Sec. \ref{Vortex}), that even
if the largest structures of the shear layer where removed by the
annulus, a pair of coupled vortices appeared. These vortices are
smaller than those we observe without the annulus, but their long
time coherent precessing must induce long time correlations that
slow down the decrease of $\bar\delta(N)$.

\bibliographystyle{unsrt}

\end{document}